\documentclass{eas}
\usepackage{graphicx}

\begin{document}

\TitreGlobal{The internal kinematics of dwarf spheroidal galaxies}

\title{The internal kinematics of dwarf spheroidal galaxies}
\author{Mark I. Wilkinson}\address{Institute of Astronomy, Madingley Road, Cambridge CB3 0HA, UK} 
\author{Jan T. Kleyna}\address{Institute for Astronomy, 2680 Woodlawn Drive,
Honolulu, Hawaii 96822-1897, USA} 
\author{N. Wyn Evans$^{1}$}
\author{Gerard F. Gilmore$^{1}$}
\author{Justin I. Read$^{1}$}
\author{Andreas Koch}\address{Astronomical Institute of the University of Basel, Venusstrasse 7,CH 4102 Binningen, Switzerland}
\author{Eva K. Grebel$^{3}$}
\author{Michael J. Irwin$^{1}$}
\runningtitle{Internal kinematics of dSphs}
\setcounter{page}{1}
\index{Wilkinson, M. I.}

\begin{abstract} 
The status of kinematic observations in Local Group dwarf spheroidal
galaxies (dSphs) is reviewed. Various approaches to the dynamical
modelling of these data are discussed and some general features of
dSph dark matter haloes based on simple mass models are presented.
\end{abstract}

\maketitle

\section{Introduction}

The dwarf spheroidal galaxies (dSph) of the Local Group constitute a
valuable testing ground for theories of dark matter and galaxy
formation. The internal kinematics of dSphs suggest that they are dark
matter dominated at all radii, making them the smallest stellar
systems known to contain dynamically significant quantities of dark
matter. Cosmological simulations of galaxy formation predict the
existence of many sub-haloes in the halo of a Milky Way type galaxy -
however, it is currently unclear which of the objects in the
simulations correspond to the observed dSphs.  A determination of the
kinematic properties of dSphs, and in particular of their masses and
mass distributions, is an important step towards understanding both
their origin in a cosmological context and their subsequent evolution
within the Local Group.

\section{Kinematic Observations in dSphs}
\begin{figure}[t]
   \centering \includegraphics[width=\textwidth]{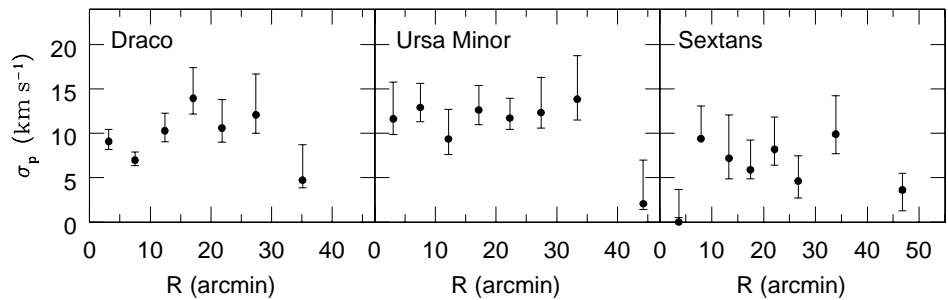}
   \caption{Velocity dispersion profiles for the Draco, Ursa Minor and
   Sextans dSphs (Kleyna et al. 2004; Wilkinson et al. 2004).}
\label{fig:disp_profiles} \end{figure}

The field of dSph kinematics had its origins in a seminal paper by
Aaronson et al. (1983) which estimated the velocity dispersion of
Draco using velocity measurements of three carbon stars. In that
paper, a mass to light ratio of $30$ was tentatively inferred from the
measured dispersion of $6.5$km\,s$^{-1}$. This provided the first hint
that the dSphs were a class of stellar system distinct from the
globular clusters. By 1987, the number of stars with measured
velocities in Draco had been increased to eleven, and an additional
ten stars had been observed in Ursa Minor (Aaronson \& Olszewski
1987). Based on the associated estimates of the velocity dispersions
of both galaxies, Pryor \& Kormendy (1990) used a family of
two-component (luminous plus dark matter) King models to demonstrate
that the central dark matter density in these dSphs must be greater
than $0.05$ M$_\odot$pc$^{-3}$. They also showed that extended dark
matter haloes in dSphs could be compatible with both the kinematics
and stellar density distributions.

Throughout the 1990s, the number of dSphs with measured velocity
dispersions increased steadily - Mateo (1998) cites dispersion values
for ten systems. However, dSph velocity distributions were still
represented by a single number, the value of the central velocity
dispersion. Mateo (1997) published the first velocity dispersion
profile for a dSph, showing the variation of the velocity dispersion
as a function of projected radius in Fornax. The measured profile,
based on 215 individual velocities, showed that the dispersion of
Fornax remains approximately flat almost to the edge of the light
distribution. This profile is inconsistent with simple models of
Fornax with a constant mass to light ratio and an isotropic velocity
distribution.

The availability of multi-fibre and multi-slit spectrographs on 4m and
10m class telescopes has revolutionised the study of dSph kinematics
by providing an efficient means to obtain large data sets of stars in
each target dSph. Kleyna et al. (2001) presented the dispersion
profile of Draco based on velocities for 159 stars obtained using the
WYFFOS multi-fibre spectrograph on the William Herschel Telescope, La
Palma. Subsequently, the measurement of dispersion profiles has
continued apace using instruments on all the major telescopes: WYFFOS
(WHT, La Palma), FLAMES/GIRAFFE \& UVES (VLT), GMOS (Gemini), MIKE
(Magellan), DEIMOS and HIRES (Keck) and dispersion profiles are now
available (or will appear shortly) for almost all the Local Group
dSphs (e.g. Kleyna et al. 2004; Tolstoy et al. 2004; Wilkinson et
al. 2004; Mu\~noz et al. 2005; Wang et
al. 2005). Figure~\ref{fig:disp_profiles} shows the profiles of Draco,
Ursa Minor and Sextans based on the WHT observations of Wilkinson et
al.(2004) and Kleyna et al. (2004).

All the teams involved in this effort measure the velocities of giant
branch stars in the dSphs, typically using the Ca triplet absorption
lines. Velocity errors of $1-5$ km\,s$^{-1}$ are routinely
obtained. In each target galaxy, the goal is to obtain a data set of
several hundred velocities covering the entire face of the stellar
distribution. In addition, higher resolution spectra (such as those
from FLAMES at the VLT) are being used to obtain abundance estimates
for large numbers of stars (e.g. 401 stars in Sculptor: Tolstoy et
al. 2004; 437 stars in Carina: Koch et al. 2005).

The new instruments and larger data sets have made it possible to
investigate empirically the extent to which the measured velocity
dispersions might be inflated by non-dynamical effects. In particular,
atmospheric ``jitter'' at the level of a few km\,s$^{-1}$, which is
observable in AGB stars, is not a problem for the stars below the tip
of the giant branch which now represent the vast majority of the
kinematic samples. The orbital velocities of binary stars were also a
serious concern for early samples based on small, single-epoch data
sets. However, Olszewski et al. (1996) used repeated observations of a
sample of 118 stars in Draco and Ursa Minor to demonstrate that the
scatter in velocities due to binaries is small compared to the
observed dispersions of the dSphs. This was later confirmed by Kleyna
et al. (2002) based on repeat observations of 61 stars in Draco from
the sample of Armandroff et al. (1995) with a baseline of 6-8
years. Thus, the large amplitudes of the observed velocity dispersions
appear to be reliable.

\subsection{Cold outer populations: are they real?}
\begin{figure}[t]
   \centering \includegraphics[width=9cm]{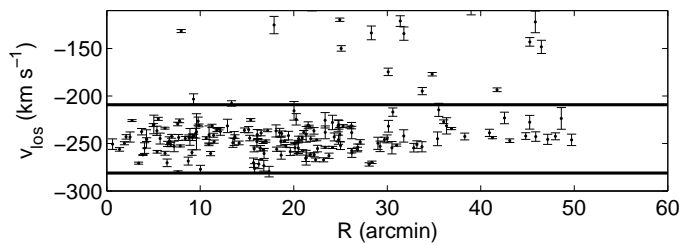} \caption{Line of
   sight velocity versus projected radius for the Ursa Minor data set
   of Wilkinson et al. (2004), including some non-members. The
   horizontal lines indicate the 3$\sigma$ velocity cuts used to
   determine membership. See text for a discussion.}
\label{fig:UMi_vels} \end{figure}

The velocity dispersion profiles of Draco and Ursa Minor shown in
Figure~\ref{fig:disp_profiles} display the unexpected property that at
large radii there is a sudden fall-off in the projected
dispersion. Such a feature is difficult to understand in the context
of smooth, equilibrium models of dSphs. Wilkinson et al. (2004)
suggest that tidal sculpting of the outer regions of a dSph by the
Milky Way could give rise to such a feature. However, Read et
al. (2005) have demonstrated that tidal effects cannot produce steeply
falling dispersion profiles, even in projection.

Recently, Mu\~noz et al. (2005) have measured velocities at very large
angular distances from Ursa Minor. Based on these velocities, they
conclude that the dispersion at the largest radii is approximately
flat. Due to the small number of stars at large radii, however, their
final dispersion point contains stars ranging in radius from
$1.5-6$kpc (corresponding to $1-4$ times the limiting radius of Ursa
Minor). Velocity gradients (such as might be expected if the stars are
associated with a tidal tail) could lead to an apparent increase in
the measured dispersion over this radial range. It is also pertinent
to consider the distribution of velocities in Ursa Minor
(Figure~\ref{fig:UMi_vels}). By eye, the velocity distribution appears
to narrow outside $37$ arcmin in this plot - $9$ out of the $12$ stars
in this region lie within error at the mean velocity of the entire
sample. Application of an F-test (e.g. Press et al. 1992) to the
velocity distributions inside and outside $37^\prime$ shows that the
outer stars have a dispersion which is $2.6$ times smaller than the
value for stars inside this radius ($90$ per cent confidence). It
therefore appears that the velocity distribution between $37^\prime$
and $50^\prime$ is indeed narrower than that inside $37^\prime$. If it
is confirmed that beyond $50^\prime$ the dispersion increases again,
then this might be consistent with the onset of tidal effects in that
region. What is clear is that the velocity distributions in the outer
regions of some dSphs are significantly more complicated than was
previously thought and further data are required to resolve the
situation. The origin of the dispersion profile features awaits
satisfactory explanation - observations are underway to increase the
numbers of observed stars at the largest radii in these dSphs.

\section{Dynamical models of dSphs}

As was stated earlier, the key goal for dynamical models of dSphs is
to determine the masses, extents and mass profiles of their dark
matter haloes. Knowledge of these properties would make it possible to
investigate, for example, the extent to which dSphs share similar halo
properties despite the wide variation in the properties of their
stellar components (stellar mass, number of distinct stellar
populations, etc). An important question which arises in the context
of mass modelling is the extent to which the dSphs have been affected
by their proximity to the Milky Way or M31. It is clear that at some
level the dSphs must have been perturbed by tidal forces. However, there
is evidence that for at least some of the dSphs, tidal effects have
not had a significant impact in the region probed by the stellar
distribution (e.g. Draco: Klessen et al. 2003). The issue of tides
remains controversial, but it is worth noting that so far no
unambiguous kinematic signature of tidal effects in a dSph has been
observed. In addition to testing models of tidal perturbation, the
dSphs also provide a potential test of MOdified Newtonian Dynamics - a
discussion of the performance of MOND in the case of Draco is
presented by {\L}okas et al. (2005).

All mass modelling of dSphs to date has been based on the assumption
of virial equilibrium. Aside from the question of whether this
assumption is valid, there are considerable uncertainties associated
with the outer surface brightness profiles of dSphs - deeper imaging
often reveals breaks in the outer profiles (e.g. Carina: Majewski et
al. 2005). As was discussed earlier, the detailed structure of the
observed outer velocity dispersion profiles is also controversial. The
uncertainties associated with the two main observed inputs necessarily
introduce large uncertainties into the resulting mass estimates which
are difficult to quantify.

\begin{figure*}[t]
\begin{minipage}{\textwidth}
   \includegraphics[width=0.51\textwidth]{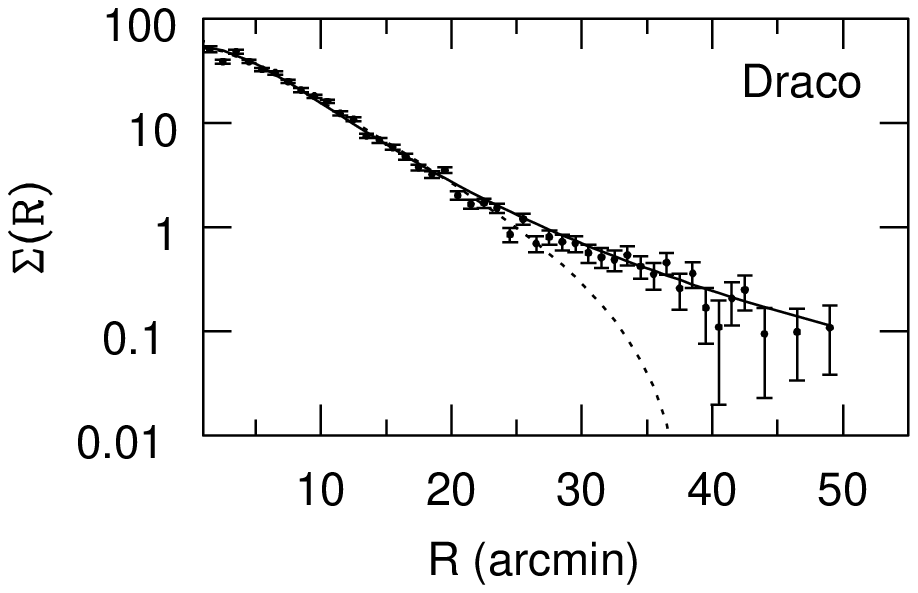}
   \includegraphics[width=0.49\textwidth]{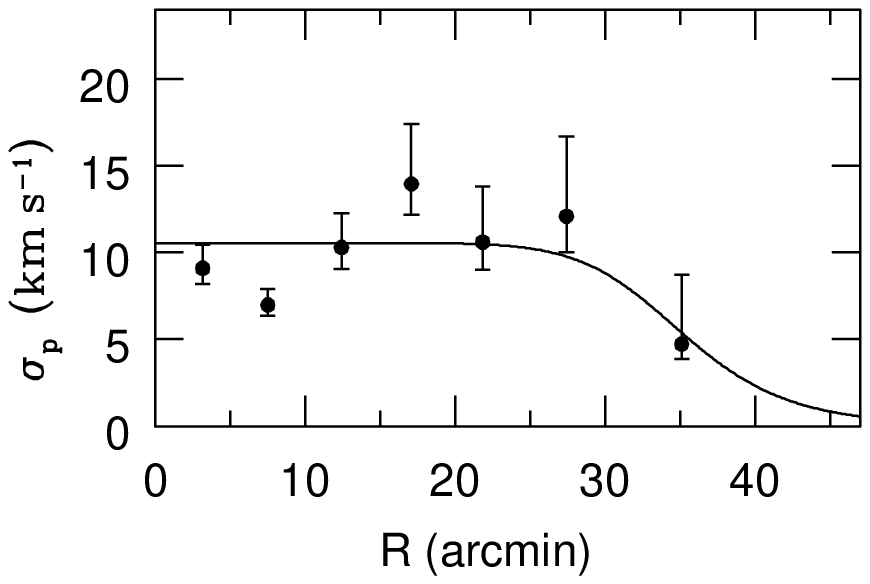}
\end{minipage}
\caption{Surface brightness and velocity dispersion profiles of Draco
(Wilkinson et al. 2004). Solid curves show the assumed smooth profiles
used in the analysis presented here. }
\label{fig:draco_models}
\end{figure*}

Notwithstanding the uncertainties in the outermost regions, a number
of authors have published mass models for dSphs based on the
(presumably) more robust data inside the break radii in the light
distributions. Pryor \& Kormendy (1990) introduced two-component
models to the study of dSphs, to explore the possibility the dSphs
have extended haloes similar to those found in external galaxies. More
recently, mass models based either on velocity moments (e.g. {\L}okas
2001 and {\L}okas et al. 2005) or on direct comparison of two
parameter distribution functions with the observed data (e.g. Kleyna
et al. 2002) have appeared in the literature. To date, Draco and Ursa
Minor have received the most attention - however, Wang et al. (2005)
have recently applied a non-parametric modelling scheme to new data on
Fornax to derive simultaneously the velocity dispersion profile and
mass profile in the context of an isotropic velocity distribution.

\section{Mass models from Jeans equations}
\begin{figure*}[t]
\begin{minipage}{\textwidth}
   \includegraphics[width=0.5\textwidth]{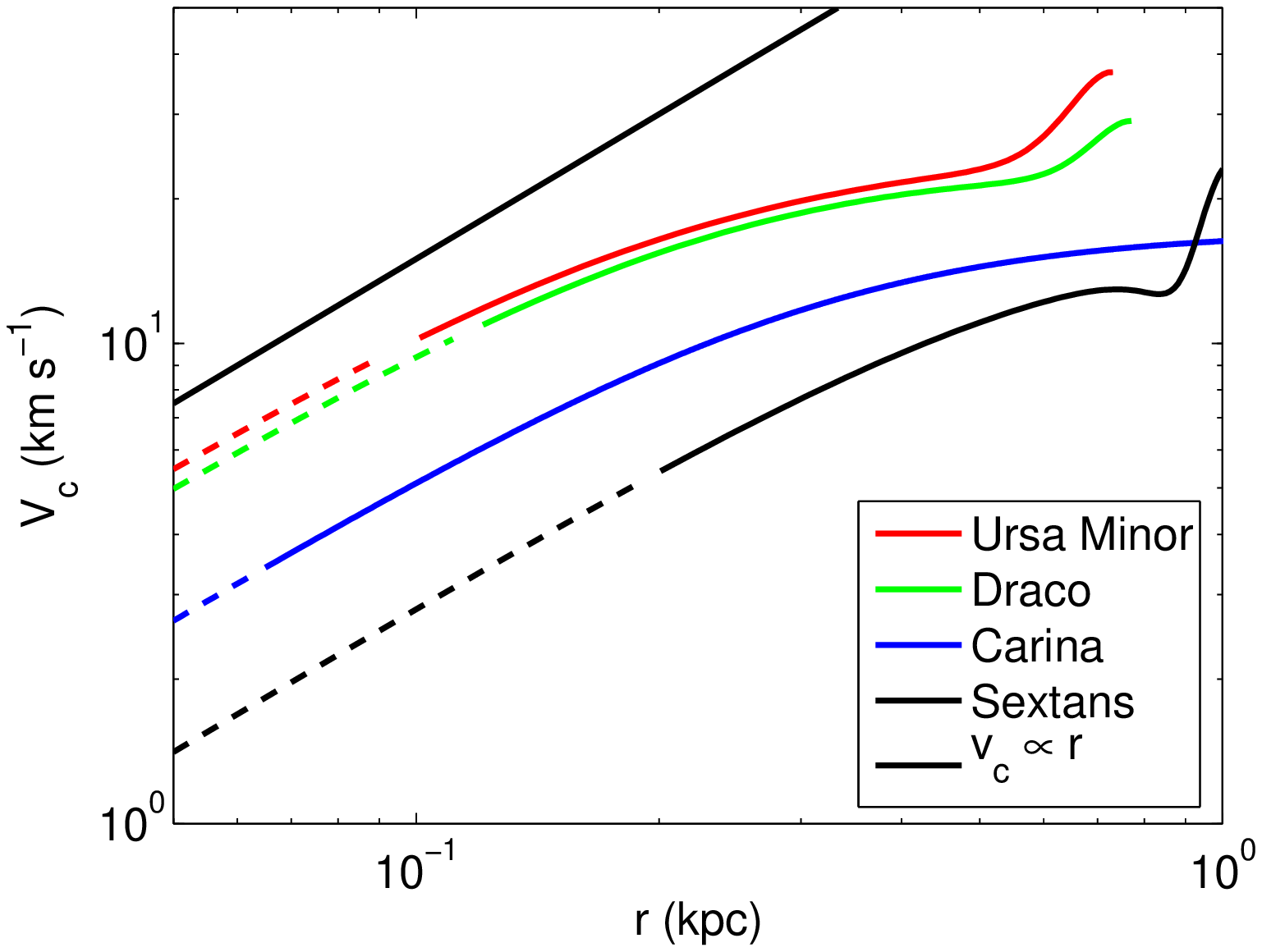}
   \includegraphics[width=0.5\textwidth]{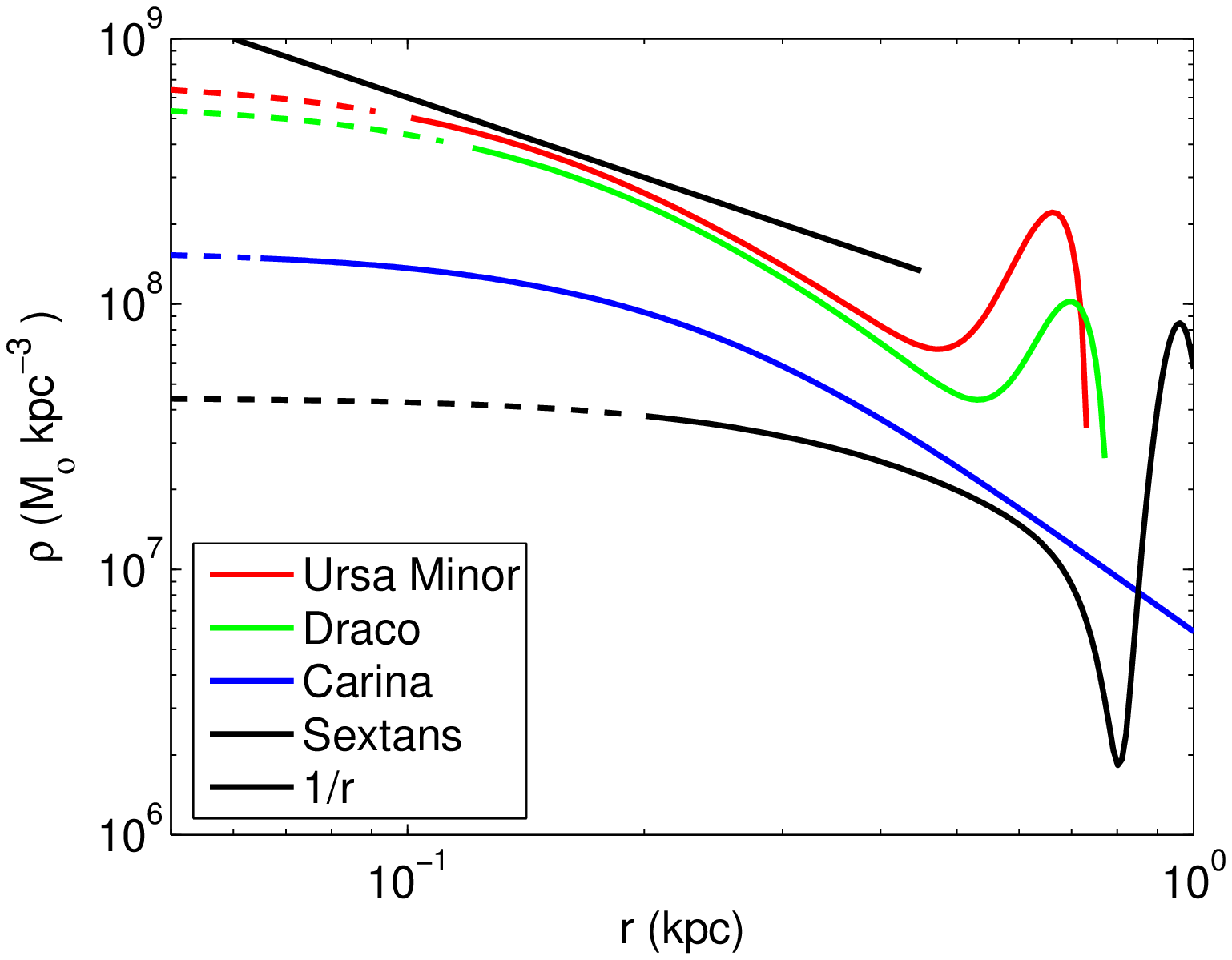}
\end{minipage}
\caption{Circular speed curves (left) and density profiles (right) for
four Milky Way dSphs, derived using Jeans equations. From top to
bottom the curves show the results for Ursa Minor, Draco, Carina and
Sextans. Also shown at the top of each plot are the expected curves
for an $r^{-1}$ density distribution. See text for a detailed
discussion.}  \label{fig:circ_dens}
\end{figure*}

To investigate the general properties of dSph haloes, we will now
derive some simple mass models for four Milky Way dSphs using Jeans
equations. We assume smooth functions to represent the light
distribution and velocity dispersion profiles of each dSph -
Figure~\ref{fig:draco_models} shows the assumed functions for
Draco. Under the simplifying assumptions of spherical symmetry and an
isotropic velocity distribution, the Jeans equations give rise to the
simple mass estimator (see e.g. Binney \& Tremaine 1987)
\begin{equation}
M(r) = -\frac{r^2}{G}\left( \frac{1}{\nu}\frac{{\rm
d}\,\nu\sigma_r^2}{{\rm d}\,r}\right)
\label{eq:jeans}
\end{equation}
where $\nu(r)$ is the three-dimensional light density distribution and
$\sigma_r(r)$ is the radial velocity dispersion. The latter two
quantities are obtained by straightforward deprojection of the surface
brightness profiles and projected velocity dispersions.

The left panel of Figure~\ref{fig:circ_dens} shows the circular speed
$v_c = \sqrt{GM(r)/r}$ as a function of radius obtained via
equation~\ref{eq:jeans}. The curves are broken in the inner regions
where either the data or the model assumptions (or both) break down -
the outer regions exhibit unphysical behaviour due to the features in
the light and/or velocity distributions which, as discussed above,
cannot be represented by simple, smooth models. The enclosed masses
(within the largest radii where the models are plausible) lie in the
range $3-8\times 10^7$M$_\odot$. The right panel of
Figure~\ref{fig:circ_dens} shows the associated mass density
distributions. Two features of these curves are noteworthy. First, all
four density profiles tend towards central logarithmic slopes which
are considerably shallower than the value of about $-1$ expected on
the basis of cosmological simulations. Secondly, more massive haloes
have higher central densities.

One should be cautious about over-interpreting the results of this
analysis, due to the simplifying assumptions which have been
made. However, on the basis of these results it is clear that the data
do not drive one to assume cusped haloes for dSphs - cored haloes with
isotropic stellar velocity distributions in the inner regions would
appear to be broadly consistent with the current data.

\begin{figure}[t]
   \centering \includegraphics[width=8cm]{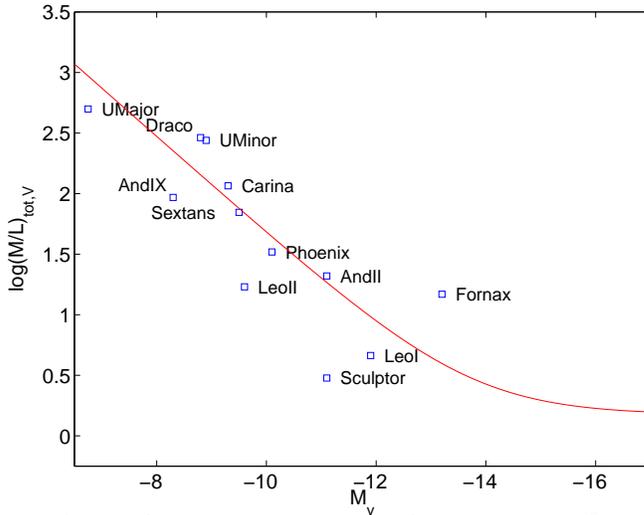} \caption{Mass
   to light ratios versus absolute magnitudes for Local Group dwarf
   galaxies. Masses for Phoenix, LeoI, LeoII and Sculptor are taken
   from Mateo et al. (1998); the mass of AndII is taken from
   C{\^o}t{\'e} et al. (1999). Recent mass estimates for the remaining
   dSphs are taken from the sources on these dSphs cited in the main
   text. See text for a discussion.}  \label{fig:ML_vs_Mv}
   \end{figure}

It is interesting to ask the question whether all the dSphs could be
embedded in dark matter haloes of similar total mass. In
Figure~\ref{fig:ML_vs_Mv}, the estimated mass to light ratios of the
local group dSphs are plotted against their V-band absolute
magnitudes. The figure is based on that given in Mateo et al. (1998)
but includes more recent mass estimates where these are available. The
solid curve shows the expected relation obtained for a population of
objects whose stellar mass to light ratio is $1.5$M$_\odot/$L$_\odot$
and whose total halo mass is $4\times 10^7$M$_\odot$. In this case,
the total mass to light ratio is simply given by $(M/L)_{\rm tot} =
(M/L)_{\rm stars} + M_{\rm dm}/L$, where $M_{\rm dm}$ is the mass in
dark matter and $L$ is the total V-band luminosity. Although there is
considerable scatter about this relation, it appears that most
observations to date are consistent with dSphs having a common halo
mass scale of around $4\times 10^7$M$_\odot$. One might legitimately
ask whether this plot contains any more information than the
similarity of dSph velocity dispersions. However, although all dSphs
seem to have velocity dispersions of $6-10$km\,s$^{-1}$, they have a
wide range of spatial extents. For a tracer population in a fixed
potential well, a larger spatial extent requires a larger velocity
dispersion to support it. Thus, this plot demonstrates that it is
possible to interpret the narrow range of dSph velocity dispersions in
terms of a common mass scale. Given that the inclusion of radius
information in the mass estimates is currently rather inhomogeneous,
it will be interesting to see how more detailed mass modelling of the
dSphs affects the details of this plot.

\section{Outlook}

As a result of the concerted efforts of a number of groups, high
quality observed velocity dispersion profiles will soon be available
for all the Milky Way dSphs. Improvements in the mass models are now
required to take advantage of the richness of these data. Of
particular interest are the correlations between kinematics and
metallicity, recently highlighted by Tolstoy et al. (2004) for the
case of Sculptor. In addition, the study of kinematic substructure in
dSphs (e.g. Kleyna et al. 2003) has the potential to yield useful
insights into the formation of dSphs. Other imminent observational
developments include the extension of surveys to the M31 system
(e.g. Chapman et al. 2005) and the search for even more dark matter
dominated systems (e.g. the Ursa Major dSph: Willman et al. 2005;
Kleyna et al. 2005).



\end{document}